\documentstyle[12pt,epsfig,oldlfont]{article}
\newcommand{\n}{\hspace*{-2.5mm}}
\newcommand{\simgt}{\rlap{\lower 3.5 pt \hbox{$\mathchar \sim$}} \raise 1pt
 \hbox {$>$}}
\newcommand{\simlt}{\rlap{\lower 3.5 pt \hbox{$\mathchar \sim$}} \raise 1pt
 \hbox {$<$}}
\newcommand{\be}{\begin{equation}}
\newcommand{\ee}{\end{equation}}

\catcode`@=11
\newcount\@tempcntc
\def\@citex[#1]#2{\if@filesw\immediate\write\@auxout{\string\citation{#2}}\fi
  \@tempcnta\z@\@tempcntb\m@ne\def\@citea{}\@cite{\@for\@citeb:=#2\do
    {\@ifundefined
       {b@\@citeb}{\@citeo\@tempcntb\m@ne\@citea\def\@citea{,}{\bf ?}\@warning
       {Citation `\@citeb' on page \thepage \space undefined}}%
    {\setbox\z@\hbox{\global\@tempcntc0\csname b@\@citeb\endcsname\relax}%
     \ifnum\@tempcntc=\z@ \@citeo\@tempcntb\m@ne
       \@citea\def\@citea{,}\hbox{\csname b@\@citeb\endcsname}%
     \else
      \advance\@tempcntb\@ne
      \ifnum\@tempcntb=\@tempcntc
      \else\advance\@tempcntb\m@ne\@citeo
      \@tempcnta\@tempcntc\@tempcntb\@tempcntc\fi\fi}}\@citeo}{#1}}
\def\@citeo{\ifnum\@tempcnta>\@tempcntb\else\@citea\def\@citea{,}%
  \ifnum\@tempcnta=\@tempcntb\the\@tempcnta\else
   {\advance\@tempcnta\@ne\ifnum\@tempcnta=\@tempcntb \else \def\@citea{--}\fi
    \advance\@tempcnta\m@ne\the\@tempcnta\@citea\the\@tempcntb}\fi\fi}
\catcode`@=12
\begin{document}
\title{\vskip-3cm{\baselineskip14pt
\centerline{\normalsize DESY 94-124\hfill ISSN 0418-9833}
\centerline{\normalsize hep-ph/9407347\hfill}
\centerline{\normalsize July 1994\hfill}}
\vskip1.5cm
Next-to-Leading Order Fragmentation Functions for Pions and Kaons}
\author{J. Binnewies, B.A. Kniehl, and G. Kramer\\
II. Institut f\"ur Theoretische Physik\thanks{Supported
by Bundesministerium f\"ur Forschung und Technologie, Bonn, Germany,
under Contract 05~6~HH~93P~(5)
and by EEC Program {\it Human Capital and Mobility} through Network
{\it Physics at High Energy Colliders} under Contract
CHRX-CT93-0357 (DG12 COMA).},
Universit\"at Hamburg\\
Luruper Chaussee 149, 22761 Hamburg, Germany}
\date{}
\maketitle
\begin{abstract}
We present new sets of fragmentation functions for charged pions and kaons,
both at leading and next-to-leading order.
They are fitted to TPC data taken at energy $\sqrt s=29$~GeV and describe
excellently a wealth of other $e^+e^-$ data on charged-hadron production,
ranging from $\sqrt s=5.2$~GeV way up to LEP~1 energy.
They also agree with data on the production of neutral pions and kaons,
if one makes the natural assumption that the respective fragmentation functions
are related to the charged counterparts by SU(2) symmetry.
We also list simple parameterizations of the $x$ and $Q^2$ dependence of
our results, which may be implemented conveniently in applications.
\end{abstract}
\newpage

\section{Introduction}

The Lagrangian of quantum chromodynamics contains quarks and gluons as
elementary fields.
Allowing for these particles to appear as asymptotic states, we can evaluate
scattering amplitudes perturbatively, in principle to arbitrary precision.
Of course, this picture violates the principle of confinement of colour;
experiments detect hadrons rather than quarks and gluons.
Nevertheless, this simplified computational procedure is very successful
in describing the production of jets of hadrons
at high transverse momenta ($p_T$) in scattering processes
and at high centre-of-mass (c.m.) energies ($\sqrt s$) in $e^+e^-$
annihilation.
Apparently, clustering partons in the final state yields a valid approximation,
although this does not account for any details of hadronization.
On the other hand, experiments are providing us with copious information
on inclusive production of single hadrons, which cannot be interpreted along
these lines.
In this case, we need a concept of describing how partons turn into hadrons.
In the framework of the QCD-improved parton model, this is achieved by
introducing fragmentation functions, $D_a^h(x,Q^2)$.
Their values correspond to the probability that a parton, $a$,
which is produced at short distance, $\sqrt{1/Q^2}$, fragments into a
hadron, $h$, carrying a fraction, $x$, of the momentum of $a$.
Unfortunately, it is not yet understood how these functions can be derived from
the QCD Lagrangian, in particular for hadrons with masses comparable to the QCD
asymptotic scale parameter, $\Lambda$.
However, given their $x$ dependence at some scale $Q_0^2$, the evolution with
$Q^2$ may be computed perturbatively in QCD.
This allows us to test QCD quantitatively within one experiment observing
single hadrons at different values of $p_T$ (in the case of scattering)
or taking data at different values of $\sqrt s$ (in the case of $e^+e^-$
annihilation).
Moreover, appealing to the factorization theorem, we may assume that the
$D_a^h(x,Q^2)$ functions are independent of the process in which they have
been determined, and characterize a universal property of $h$.
This enables us to make quantitative predictions for other experiments, too.

Recently, precise data on inclusive hadron production at
$p\bar p$, $ep$, and $e^+e^-$ colliders have become available.
A comprehensive review of inclusive particle spectra in earlier $e^+e^-$
experiments may be found in \cite{mattig}.
In a series of papers, two of us---partly in collaboration with
Borzumati---have
confronted $p\bar p$ data by CDF and UA2 and $ep$ data by H1 on pion and
kaon production with theoretical predictions at NLO \cite{bor,fmb}.
Although we found reasonable agreement, we have to bear in mind that
a possible limitation of our analyses is related to the point that
the literature on pion and kaon fragmentation functions is now rather aged
and, perhaps, slightly obsolete.
The most recent sets we could locate were the ones by Baier et al.\
\cite{baier}, which were updated and complemented by Anselmino et al.\
\cite{ansel}.
These sets are LO fits to low-energy $e^+e^-$ and deep-inelastic muon-nucleon
data.

Recently, NLO sets for neutral pions were presented in \cite{chiap}.
These were fitted partly to $e^+e^-$ data in the 30~GeV range,
partly to fixed-target and collider data, with
23~GeV${}\le\sqrt s\le{}$630~GeV, and partly to data of $e^+e^-$
annihilation at 30~GeV simulated using a Monte Carlo event generator.
Unfortunately, only the $x$ distributions at $Q_0^2$ are given in \cite{chiap};
an implementation of the $Q^2$ dependence is not available.
New NLO sets have also been constructed for the eta meson and charged pions
just recently \cite{greco}, again using Monte Carlo data.
Also here, the $Q^2$ dependence is not given.

Since our goal is to test quantitatively the QCD-improved parton model,
we consider it requisite to perform an independent analysis based
on real experimental information.
Complementarily to \cite{chiap},
we shall focus attention on charged pions and kaons, averaged separately.
Similarly to \cite{chiap}, we analyze $e^+e^-$ data at around 30~GeV.
A different avenue was taken in \cite{nason}, where charged-hadron
fragmentation functions were extracted from LEP~1 data, not distinguishing
between the various particle species.

The organization of this paper is as follows.
In Sect.~2, we shall introduce the formalism needed to extract fragmentation
functions from $e^+e^-$ data on inclusive hadron production at NLO.
In Sect.~3, we shall carry out the actual analysis and discuss our results.
In Sect.~4, we shall check our results against data on neutral-pion
and -kaon production as well as data on charged-hadron production without
separation of the particle species.
We shall also compare our results with those of \cite{nason}.
Our conclusions will be summarized in Sect.~5.
In the Appendix, we shall list simple parameterizations of our sets and quote
their range of validity.

\section{Formalism}

The inclusive production of a hadron, $h$, by $e^+e^-$ annihilation,
\be
\label{process}
e^+e^-\to(\gamma,Z)\to h+X,
\ee
is completely characterized by three observables.
These are the energy fraction of the outgoing hadron,
\be
\label{defx}
x=\frac{2E_h}{\sqrt{s}},
\ee
its angle with the beam axis, $\theta$, and the total energy of the system,
$\sqrt s$.
For massless hadrons, definition~(\ref{defx}) agrees with the convention
used in experiments, whereas for nonvanishing masses, $m_h$, one has to adjust
the theoretical prediction by including a factor
$\beta_h=\sqrt{1-m_h^2/E_h^2}$.
Choice of reference frame is not an issue here,
since, at $e^+e^-$~colliders, the laboratory and c.m.\ frames coincide.
The cross section of (\ref{process}) exhibits the angular structure
\be
\label{sigxtheta}
\frac{d^2\sigma(e^+e^-\to h+X)}{dx\,d{\cos}\theta}=\frac{3}{8}
     (1+{\cos}^2\theta)\frac{d\sigma^T}{dx}
     +\frac{3}{4}{\sin}^2\theta\frac{d\sigma^L}{dx}
     +\frac{3}{4}{\cos}\theta\frac{d\sigma^A}{dx},
\ee
where the superscripts $T$ and $L$ denote the contributions due to
transverse and longitudinal polarizations, respectively, and the asymmetric
term, labelled $A$, accounts for the interference of the photon with the $Z$
boson.
Usually, experiments do not determine $\theta$ distributions.
Thus, we shall integrate over $\theta$.
This eliminates the asymmetric term in (\ref{sigxtheta}).

The partonic subprocesses may be treated exactly in perturbative QCD.
However, this is not yet possible for the process of hadronization.
As explained in the Introduction, the latter is described in terms
of phenomenological fragmentation functions, which must be extracted
from experiment.
In the language of the QCD-improved parton model, the $x$ distribution of
(\ref{process}) emerges from the $x$ distribution of $e^+e^-\to a+X$,
$(d\sigma_a/dx)(x,\mu^2,Q^2)$, through
convolution with $D_a^h(x,Q^2)$,
\be
\label{sigx}
\frac{s}{\beta_h}\,\frac{d\sigma(e^+e^-\to h+X)}{dx}=\sum_a
\int\limits_x^1\frac{dz}{z}\,D_a^{h}(z,M_f^2)\,s
\frac{d\sigma_a}{dy}\left(\frac{x}{z},\mu^2,M_f^2\right).
\ee
Here, the sum extents over all active partons ($a=g,u,d,s,c,b$),
$\mu$ is the renormalization scale of the partonic subprocess,
and $M_f$ is the so-called fragmentation scale,
which corresponds to the typical energy where the transition from the
perturbative to the non-perturbative regime takes place.
In next-to-leading order (NLO), $M_f$ defines the point where the
divergence associated with collinear radiation off parton $a$ is to be
subtracted.
As usual, we work in the massless-quark approximation, i.e., we neglect
threshold effects.
As a consequence, the bottom quark is taken to be active throughout the
entire $Q^2$ evolution and to contribute to the cross section also at
low energies.
We checked that this slight inconsistency is inconsequential for the goodness
of our fits.
To NLO in the $\overline{\rm MS}$ scheme, the cross sections of the
relevant subprocesses are given by~\cite{soyoungpi}
\begin{eqnarray}
\label{xsection}
\frac{d\sigma_{q_i}}{dy}(y,\mu^2,M_f^2)&\n=\n&
\sigma_0N_ce_{q_i}^2\left[\delta (1-y)+\frac{\alpha_S(\mu^2)}
{2\pi}\left(P_{qq}^{(0,T)}(y)\ln{\frac{s}{M_f^2}}
+K_q^T(y)+K_q^L(y)\right)\right],\nonumber\\
\frac{d\sigma_g}{dy}(y,\mu^2,M_f^2)&\n=\n&
\sigma_0N_c\left(\sum_{i=1}^{2N_f}e_{q_i}^2\right)
\frac{\alpha_S(\mu^2)}{2\pi}\left(P_{qg}^{(0,T)}(y)\ln{\frac{s}{M_f^2}}
+K_g^T(y)+K_g^L(y)\right),
\end{eqnarray}
where $\sigma_0=(4\pi\alpha_{\rm QED}^2/3s)$ is the total cross
section of $e^+e^-\to\mu^+\mu^-$ for massless leptons, $N_c=3$,
$e_{q_i}$ is the electric charge of quark $q_i$ in units of the positron
charge, $P_{ab}^{(0,T)}(y)$ is the LO term of the transverse $a\to b$
splitting function,
\be
P_{ab}^T(y,\alpha_S(Q^2))=P_{ab}^{(0,T)}(y)+\frac{\alpha_S(Q^2)}{2\pi}
P_{ab}^{(1,T)}(y)+\cdots,
\ee
and
\begin{eqnarray}
K_q^T(y)&\n=\n&
C_F\left[\frac{3}{2}(1-y)-\frac{3}{2}\,\left(\frac{1}{1-y}\right)_+
+2\frac{1+y^2}{1-y}\ln{y}\right.\nonumber\\
& &{}+\left.(1+y^2)\left(\frac{\ln{(1-y)}}{1-y}\right)_+
+\left(\frac{2\pi^2}{3}-\frac{9}{2}\right)\delta(1-y)\right],\nonumber\\
K_q^L(y)&\n=\n&C_F,\nonumber\\
K_g^T(y)&\n=\n&C_F\left[\frac{1+(1-y)^2}{y}\Bigl(\ln{(1-y)}+2\ln{y}\Bigr)
-2\frac{1-y}{y}\right],\nonumber\\
K_g^L(y)&\n=\n&2C_F\frac{1-y}{y},
\end{eqnarray}
with $C_F=(N_c^2-1)/(2N_c)=4/3$.
As usual, the subscript + denotes distributions.
The NLO formula of $\alpha_s(\mu)$ may be found, e.g., in \cite{pdg}.
Throughout this work, we choose $\Lambda=190$~MeV, in order to be compatible
with recent sets of parton density functions for the proton and photon.
The scales $\mu$ and $M_f$ are identified with the only intrinsic
energy scale, $\sqrt{s}$. For this choice of scales, the logarithmic terms in
(\ref{xsection}) involving $s$ vanish, so that the NLO corrections are
expressed just in terms of the $K$ functions.
We observe that $K_g^T(y)+K_g^L(y)<0$ for any value of $y$,
so that the gluon contributes destructively to (\ref{sigx}).

Having defined the partonic subprocesses, we turn to the
fragmentation functions.
Their $x$ distributions are not yet calculable in the framework of
perturbative QCD.
However, once we know them at some scale $Q_0^2$,
the $Q^2$ evolution is determined by the DGLAP equations~\cite{ap1}.
Our task is thus to construct a model for the $x$ distributions at a
starting scale $Q_0^2$, which, after evolution, fits the data at scale $Q^2$.
The DGLAP equations read
\be
\frac{d}{d\ln{Q^2}}D_a^h(x,Q^2)
=\frac{\alpha_S(Q^2)}{2\pi}\sum_{b}\int_x^1{dy\over y}
P_{ba}^T(y,\alpha_S(Q^2))D_{b}^h\left({x\over y},Q^2\right).
\ee
That is, we have to solve a system of integro-differential equations.
This may be achieved with the help of the Mellin-transform technique
\cite{furmanski}.
The Mellin transform, $F$, of a function, $f$, is defined as the set of
its moments,
\be
F(n)=\int\limits_0^1dx\,x^{n-1}f(x).
\ee
The inverse transformation is
\be
\label{mellinfro}
f(x)=\frac{1}{2\pi i}\int\limits_{\gamma-i\infty}^{\gamma+i\infty}dn\,
x^{-n}F(n),
\ee
where the path of integration must be chosen to the right of all
singularities of the integrand.
In our numerical analysis, we choose $\gamma=10$.
The essential property of this transformation is that it renders
convolutions to products.
In fact, denoting the moments of $P_{ab}^T$ and $D_a^h$ by
$A_{ab}$ and $M_a$, respectively, we have
\begin{eqnarray}
\label{glapn}
\frac{d}{d\ln{Q^2}}M_i^-(n,Q^2)&\n=\n&\frac{\alpha_S(Q^2)}{2\pi}
A_{NS}(n,\alpha_S(Q^2))M_i^-(n,Q^2),\nonumber\\
\frac{d}{d\ln{Q^2}}M_i^+(n,Q^2)&\n=\n& \frac{\alpha_S(Q^2)}{2\pi}
A_{NS}(n,\alpha_S(Q^2))M_i^+(n,Q^2),\nonumber\\
\frac{d}{d\ln{Q^2}}M_{\Sigma}(n,Q^2)&\n=\n& \frac{\alpha_S(Q^2)}{2\pi}
\left(A_{qq}(n,\alpha_S(Q^2))M_{\Sigma}(n,Q^2)
+A_{gq}(n,\alpha_S(Q^2))M_G(n,Q^2)\right),\nonumber\\
\frac{d}{d\ln{Q^2}}M_G(n,Q^2)&\n=\n& \frac{\alpha_S(Q^2)}{2\pi}
\left(A_{qg}(n,\alpha_S(Q^2))M_{\Sigma}(n,Q^2)
+A_{gg}(n,\alpha_S(Q^2))M_G(n,Q^2)\right),\qquad
\end{eqnarray}
where $A_{NS}$ refers to the usual non-singlet combination of quark and
antiquark splitting functions \cite{furmanski}, and
\begin{eqnarray}
M_i^-(n,Q^2)&\n=\n&\frac{1}{2}\left(M_{q_i}(n,Q^2)
        -M_{\bar{q}_i}(n,Q^2)\right),\nonumber\\
M_i^+(n,Q^2)&\n=\n&\frac{1}{2}\left(M_{q_i}(n,Q^2)
        +M_{\bar{q}_i}(n,Q^2)\right)
        -\frac{1}{2N_f}M_{\Sigma}(n,Q^2),\nonumber\\
M_{\Sigma}(n,Q^2)&\n=\n&\sum_{i=1}^{N_f}
        \left(M_{q_i}(n,Q^2)
        +M_{\bar{q}_i}(n,Q^2)\right),\nonumber\\
M_G(n,Q^2)&\n=\n& M_g(n,Q^2).
\end{eqnarray}
The LO expressions for $A_{ab}$ may be found in
\cite{furmanski,floratos}\footnote{%
Note that there is a typographical error in (B.46c) of \cite{floratos}.
$C_F$ should be replaced by $C_A$.}
and the NLO expressions in \cite{grv2}\footnote{%
Note that there is a typographical error in (A.10) of \cite{grv2}.
The term before the last one of the contribution proportional to $C_FT_Rf$
should read $-(16/3(n+2)^2)$ instead of $-(16/3(n+1)^2)$.}.
Note that the non-singlet terms, $M_i^{\pm}$, decouple from the gluon.
The solutions of (\ref{glapn}) factorize,
\be
\label{fragevn}
M(n,Q^2)=E(Q_0^2,Q^2)\,M(n,Q_0^2).
\ee
The relevant evolution operators, $E(Q_0^2,Q^2)$, may be found in
\cite{furmanski}.
We parameterize the $x$ dependence of the fragmentation functions at $Q_0^2$ as
\be
\label{frag0x}
D(x,Q_0^2)=Nx^{\alpha}(1-x)^{\beta}.
\ee
The corresponding moments read
\be
M(n,Q_0^2)=NB(n+\alpha,\beta+1),
\ee
where $B$ is Euler's beta function.
The evolution in $n$ space is performed analytically via (\ref{fragevn}).
Finally, the $D(x,Q^2)$ functions are obtained by solving (\ref{mellinfro})
numerically.
This requires knowledge of $A_{ab}$ for complex $n$.
A prescription for the proper analytic continuation may be found in
\cite{grv1}.

In (\ref{frag0x}), we introduced three parameters for each parton.
To reduce the number of free parameters, we shall discriminate only between
valence- and sea-type quarks.
To that end, we make the following ansatz for the average of charged pions,
\begin{eqnarray}
D_V^\pi+D_S^\pi&\n=\n&
D_u^\pi=D_{\bar u}^\pi=D_d^\pi=D_{\bar d}^\pi,\nonumber\\
D_S^\pi&\n=\n&
D_s^\pi=D_{\bar s}^\pi=D_c^\pi=D_{\bar c}^\pi=D_b^\pi=D_{\bar b}^\pi.
\end{eqnarray}
For the average of the charged kaons, we set
\begin{eqnarray}
\label{kansatz}
D_V^K+D_S^K&\n=\n&
D_u^K=D_{\bar u}^K=D_s^K=D_{\bar s}^K,\nonumber\\
D_S^K&\n=\n&
D_d^K=D_{\bar d}^K=D_c^K=D_{\bar c}^K=D_b^K=D_{\bar b}^K.
\end{eqnarray}
Furthermore, we put $\alpha=-1$ for valence and sea quarks,
which may be motivated by inspection of the energy dependence of hadronic
multiplicities.
In the case of the gluon, we found that the choice $\alpha=0$ is favoured
by the data.
We are thus left with six parameters to be determined from experiment.

\section{Results}

In general, data of inclusive hadron production at $e^+e^-$ colliders
are most suitable for the extraction of fragmentation functions;
in the case of fixed-target, collider, and $ep$ data, the information
on the fragmentation functions is obscured by theoretical uncertainties
arising from the parton density functions and the choice of factorization
scales connected with the initial state.
For our analysis, we select the data on charged-pion and -kaon production
taken at energy $\sqrt{s}=29$~GeV by the TPC Collaboration at SLAC \cite{tpc2}.
These data combine small statistical errors with fine binning in $x$
and are thus more constraining than data collected by other
$e^+e^-$ experiments in the energy range from 5.2 to 44~GeV.
For this reason, our approach is to fit exclusively to TPC data and to use
the other data for cross checks.

The quality of the fit is measured in terms of the average
$\chi^2_{\rm d.o.f.}$ for all data points with
$x>x_{\rm cut}=2\mbox{ GeV}/\sqrt{s}$. This
restriction is necessary to exclude events in the
non-perturbative region, where the above formalism is bound to fail.
Our technical procedure is as follows.
We consider (\ref{sigx}) at $\mu=M_f=\sqrt s=29$~GeV as a function of the
six parameters that determine the $x$ dependence of the fragmentation
functions at $Q_0^2=2$~GeV$^2$.
Using a multidimensional minimization algorithm \cite{minuit},
we search this six-dimensional parameter space for the point at which
the deviation of the theoretical prediction from the data becomes minimal.
Our results are listed below.
For the average of charged pions, we obtain
\begin{eqnarray}
\label{pilo}
D_V^{(\pi,LO)}(x,Q_0)&\n=\n&0.551\,x^{-1}\,(1-x)^{1.2},\nonumber\\
D_S^{(\pi,LO)}(x,Q_0)&\n=\n&1.23\,x^{-1}\,(1-x)^{4.77},\nonumber\\
D_G^{(\pi,LO)}(x,Q_0)&\n=\n&3.77\,(1-x)^{2.0}
\end{eqnarray}
in LO and
\begin{eqnarray}
\label{pinlo}
D_V^{(\pi,NLO)}(x,Q_0)&\n=\n&0.338\,x^{-1}\,(1-x)^{1.2},\nonumber\\
D_S^{(\pi,NLO)}(x,Q_0)&\n=\n&1.19\,x^{-1}\,(1-x)^{5.26},\nonumber\\
D_G^{(\pi,NLO)}(x,Q_0)&\n=\n&3.65\,(1-x)^{2.0}
\end{eqnarray}
in NLO.
In the case of kaons, we find
\begin{eqnarray}
D_V^{(K,LO)}(x,Q_0 )&\n=\n&0.155\,x^{-1}\,(1-x)^{1.4},\nonumber\\
D_S^{(K,LO)}(x,Q_0 )&\n=\n&0.216\,x^{-1}\,(1-x)^{1.69},\nonumber\\
D_G^{(K,LO)}(x,Q_0 )&\n=\n&0.324\,(1-x)^{2.0}
\end{eqnarray}
in LO and
\begin{eqnarray}
\label{kanlo}
D_V^{(K,NLO)}(x,Q_0 )&\n=\n&0.099\,x^{-1}\,(1-x)^{1.4},\nonumber\\
D_S^{(K,NLO)}(x,Q_0 )&\n=\n&0.214\,x^{-1}\,(1-x)^{2.16},\nonumber\\
D_G^{(K,NLO)}(x,Q_0 )&\n=\n&0.372\,(1-x)^{2.0}
\end{eqnarray}
in NLO.
For the reader's convenience, we list simple parameterizations of the $x$ and
$Q^2$ dependences of these sets in the Appendix.
It cannot be expected that such simple parameterizations reproduce the true
result perfectly for all $x$ and $Q^2$ values that occur in applications.
Deviations in excess of 10\% may occur at $Q^2<25$~GeV$^2$ and $x<0.1$,
in particular for the gluon.
Nevertheless, we believe that such parameterizations are indispensable for
practical purposes, especially at NLO.

Our fragmentation functions exhibit good perturbative stability.
In the case of the pions, there is only little difference between our
LO and NLO sets.
The difference is somewhat bigger for the kaons.
In the comparisons with data, the NLO sets do just slightly better.
The $x$ dependence of (\ref{pinlo},\ref{kanlo}) at $Q_0^2$ is visualized in
Figs.~\ref{fig1},\ref{fig5}, respectively.
In the following, we shall always employ the parameterizations given in the
Appendix.
The goodness of our fits to the TPC data on charged-pion and -kaon production
may be judged from Figs.~\ref{fig2},\ref{fig6}, respectively.
In LO (NLO), we achieve $\chi^2_{\rm d.o.f.}=0.7$ (0.7) for pions
and 0.6 (0.5) for kaons.
Here and in the following, $\chi^2_{\rm d.o.f.}$ refers to the interval
$\max(x_{\rm cut},0.1)\le x\le0.9$.
Notice that our parameterizations fit the TPC data excellently even down to
$x=0.06$.

In Figs.~\ref{fig3}--\ref{fig7}, we postdict data taken by DASP \cite{dasp}
and ARGUS \cite{argus} at $\sqrt s=5.2$ and 10~GeV, respectively.
We have made comparisons also with TASSO data at $\sqrt s=34$ and 44~GeV
\cite{tasso2}, which we do not, however, display graphically.
Data of inclusive charged-kaon production are well reproduced by our fits.
In Fig.~\ref{fig7}, this is nicely demonstrated for ARGUS.
In the case of the charged-kaon data by DASP, for which we do not show a
figure, we reach $\chi^2_{\rm d.o.f.}=1.4$ (1.2) in LO (NLO),
which we can bring down to values below unity when we exploit the freedom
in the overall normalization of the data by about $\pm12\%$.
The charged-kaon data by the other experiments \cite{tpc2,argus,tasso2}
yield $\chi^2_{\rm d.o.f.}$ values between $0.4$ and $0.6$ without adjustment,
both in LO and NLO.

The situation is slightly less favourable in the case of charged pions.
The pion data by DASP \cite{dasp} behave similarly as their kaon data
(see Fig.~\ref{fig3}).
However, the ARGUS data \cite{argus} are inconsistent intrinsically at
large $x$, which leads to increased $\chi^2_{\rm d.o.f.}$ values,
$2.2$ in LO and $1.8$ in NLO (see Fig.~\ref{fig4}).
For $x<0.1$, the theoretical curves overshoot the data.
This is presumably attributed to mass effects, which have been ignored in our
approach.
As per construction, the TPC data are reproduced well, with
$\chi^2_{\rm d.o.f.}=0.7$ both in LO and NLO.
The TASSO data \cite{tasso2} are fairly described,
with $\chi^2_{\rm d.o.f.}$ values between $1.4$ and $2.0$.
However, these data are of limited use for our purposes because, for $x>0.2$,
their binning is a factor of 3 to 5 coarser than that of the TPC data.
Furthermore, the data sets at $\sqrt s=34$ and 44~GeV do not comply with the
expectation that scaling violation renders spectra at higher energies steeper.
Leaving aside the TASSO data,
the goodness of our charged-pion fits is quite satisfactory.

\section{Applications}

Since fragmentation functions are assumed to be universal, i.e.,
independent of the process from which they have been extracted,
we can apply them to make predictions for inclusive particle production at
fixed-target, collider, and $ep$ scattering experiments, too.
In the following, we shall confine ourselves to more extended studies of
$e^+e^-$ annihilation, which will substantiate our confidence in the
virtue of our sets.
Work on inclusive particle production in NLO at HERA and $p\bar p$ colliders
is in progress.

We commence by elaborating a welcome generalization of our results.
By SU(2) symmetry, our charged-pion sets are valid also for the neutrals.
This observation is supported by the inspection of neutral-pion data, e.g.,
by ARGUS \cite{argus0}, TPC \cite{tpc0}, JADE \cite{jade0}, and CELLO
\cite{cello0},
which agree with the respective charged-pion data within the errors quoted.
This is reflected also by the $\chi^2_{\rm d.o.f.}$ analysis.
Cross checks with new data by L3 \cite{newl3} prove to be very successful,
too.
By analogous theoretical considerations, we can establish a relationship
between the fragmentation functions for the average of the $K^0$ and
$\overline{K}^0$ mesons and those for the charged kaons.
Specifically, the $u,\bar u$ and $d,\bar d$ fragmentation functions swop
places in (\ref{kansatz}).
Again, we find reasonable agreement way up to LEP~1 energy
\cite{argus,newl3,hrs,tassok,aleph}.

As a second application, we deal with inclusive charged-hadron production.
I.e., we consider experiments that observe charged tracks without
identifying the particle species.
In this context, it has become customary to sum over the produced hadrons
rather than to form the average.
Clearly, pions and kaons make up the bulk of the cross section.
By including also protons and antiprotons, the next important subset,
we obtain a good approximation for the sum of all charged hadrons.
Inspection of TPC data \cite{tpc2} reveals that the averaged cross section
of inclusive production of protons or antiprotons has a shape similar to
the one of charged pions, but is reduced in size by a factor of $0.16\pm0.02$.
Thus, we take the hadronic cross section to be
\be
\label{sighadc}
\frac{d\sigma^{(h^++h^-)}}{dx} =
2\left[(1+0.16)\frac{d\sigma^{(\pi^+ + \pi^-)/2}}{dx}
+ \frac{d\sigma^{(K^+ + K^-)/2}}{dx}\right].
\ee
We compare this prediction with high-precision data by
DELPHI \cite{delphisig} in Fig.~\ref{fig13}
and find good agreement, with $\chi^2_{\rm d.o.f.}=0.5$ (1.0) in LO (NLO).
Our results also agree well with data by MARK~II \cite{mark2sig},
AMY \cite{amysig}, TASSO \cite{tassosig}, and CELLO \cite{cellosig}.
Whereas the $x$ dependence is reproduced perfectly for the MARK~II, AMY, and
DELPHI data, there are minor deviations in the case of TASSO and CELLO.
Except for DELPHI, the $\chi^2_{\rm d.o.f.}$ values exceed unity
in all these cases, which indicates that the incorporation of baryons
by an overall $Q^2$-independent factor is perhaps to na\"\i ve.

In order to test scaling violation, we compute (\ref{sighadc}) at
$\sqrt s=91$ and 29~GeV, plot the ratio as a function of $x$ in
Fig.~\ref{fig14}, and compare the answer with the corresponding ratio
of DELPHI \cite{delphisig} and MARK~II data \cite{mark2sig},
taking into account only the statistical errors of the latter.
For $x>0.1$, our prediction agrees remarkably well with the experimental
result, both in LO (dashed line) and NLO (solid line).
We note in passing that a similar analysis was carried out recently by Nason
and Webber \cite{nason}, who encountered a certain mismatch between their
calculation and their choice of data.
These authors determined a set of fragmentation functions by fitting to the
DELPHI data on inclusive charged-hadron production \cite{delphisig},
evolved their results backwards down to PETRA energy, and
compared the outcome with TASSO data \cite{tassosig}, taking into account
only the statistical errors of the latter.
They interpreted the discrepancy that they discovered by suggesting that
significant nonperturbative power corrections are necessary to reproduce
the measured scaling violation (see Fig.~14 of \cite{nason}).
In our opinion, the major source of this problem is the neglect of the
systematic errors on the TASSO data rather than the influence of power
corrections.
In fact, when we repeat the analysis of Fig.~\ref{fig14} for TASSO and CELLO
\cite{cellosig} data, both taken at PETRA, again including only the
statistical errors in the denominator,
we find that the two sets do not overlap and are separated by our curves.
This suggests that the TASSO and CELLO data are not consistent with each other
when only their statistical errors are taken into account.

Another look at scaling violation is presented in Fig.~\ref{fig10},
which shows the $Q^2$ dependence of the charged-hadron cross section
for various values of $x$.
The evaluations of (\ref{sighadc}) to LO (dashed lines) and NLO (solid lines)
are confronted with a rich collection of data
\cite{dasp,argus,delphisig,mark2sig,amysig,tassosig,cellosig},
which have been interpolated in $x$ where appropriate.
We conclude that the data indeed exhibit the expected scaling violation.

The DELPHI Collaboration used an alternative approach to interpret their
data \cite{delphisig}.
The inclusive hadron spectra are generated from a Monte Carlo program based
on order $\alpha_S^2$ QCD matrix elements, which produces jet cross sections
and takes into account the subsequent fragmentation of the jets into hadrons
according to certain model assumptions.
In this approach, the cancellation of infrared and collinear singularities
occurs through the definition of jet cross sections with chosen jet
resolution criteria.
This deviates from the canonical factorization approach, where the collinear
singularities are absorbed into the fragmentation functions.
We strongly advocate the canonical formalism, which is intrinsic to the
QCD-improved parton model.

At this point, we should address a possible shortcoming of our analysis,
which is common to all studies of fragmentation in $e^+e^-$ annihilation,
namely the circumstance that the gluon fragmentation function is poorly
determined.
This is due to the fact that the gluon participates in the process only at
NLO, and even then has an appreciable impact on the cross section only at
very small $x$ values close to $x_{\rm cut}$.
Thus, it contributes mainly through the $Q^2$ evolution, where it couples to
the singlet combination of quarks.
Consequently, the gluon parameters are strongly correlated with the sea-quark
parameters and are feebly constrained by the data.
Hadron and $ep$ scattering experiments deliver far more information on the
gluon fragmentation function.
Fortunately, the new three-jet data by OPAL \cite{opal} are able to probe
gluon fragmentation directly.
In Fig.~\ref{figopal}, we compare these data with our LO (dashed line) and
NLO predictions (solid line) according to (\ref{sighadc}).
The agreement is reasonable.
We tried also different ans\"atze for $D_G(x,Q_0^2)$ by choosing $\alpha\ne0$
in (\ref{frag0x}).
While they worked similarly well for all $e^+e^-$ data considered above,
they failed in the case of Fig.~\ref{figopal} \cite{dipl}.

Finally, we investigate if our fragmentation functions satisfy the momentum
sum rules.
Guided by the idea that a given outgoing parton, $a$, will fragment with
100\% likelihood into some hadron, $h$, and that momentum is conserved during
the fragmentation process, we expect that
\be
\label{sumrule}
\sum_h\int_0^1dx\,xD_a^h(x,Q^2)=1
\ee
holds for any value of $Q^2$.
In our analysis, the left-hand-side of (\ref{sumrule}) should be smaller than
unity, since we consider only pions, kaons, protons, and antiprotons,
while we do not include hyperon channels.
Furthermore, we are forced to introduce a finite lower bound of integration in
(\ref{sumrule}), since the $x\to0$ limit of $D_a^h(x,Q^2)$ is beyond our
control.
$x_{\rm min}$ is a plausible choice for this cutoff, except at small $Q^2$
($\sqrt{Q^2}<20$~GeV), where we use 0.1 instead.
In Table~I, we list the results obtained with our NLO fragmentation functions
for $\sqrt{Q^2}=5$, 29, and 91~GeV.
In fact, all entries are smaller than one as they should.

\begin{table} {TABLE~I. Left-hand side of (\ref{sumrule}) at NLO for
$\sqrt{Q^2}=5$, 29, and 91~GeV.
We use $\min(x_{\rm cut},0.1)$ as the lower bound of integration and sum over
pions, kaons, protons, and antiprotons.}\\[1ex]
\begin{tabular}{|c|c|c|c|} \hline
$a$ & \multicolumn{3}{c|}{$\sqrt{Q^2}$ [GeV]} \\
\cline{2-4} & 5 & 29 & 91 \\
\hline
$u,d$ & 0.81 & 0.76 & 0.94 \\
$s$   & 0.55 & 0.52 & 0.68 \\
$c,b$ & 0.45 & 0.45 & 0.60 \\
$g$   & 0.62 & 0.58 & 0.83 \\
\hline
\end{tabular}
\end{table}

\section{Conclusions and Outlook}

We presented LO and NLO fragmentation functions for charged pions and kaons in
the $\overline{\rm MS}$ scheme.
They were extracted from $e^+e^-$ data on inclusive production of these
particles collected at $\sqrt s=29$~GeV.
We checked that these fragmentation functions are in agreement with
the majority of the $e^+e^-$ data in the energy range between 5.2~GeV and
$M_Z$,
and suggested possible explanations, where this was not the case.
This work is motivated by the circumstance that a great amount of
high-quality data on inclusive hadron production has been accumulated in
fixed-target, hadron collider, $ep$ scattering, and $e^+e^-$ annihilation
experiments since the early 1980's, which was not accompanied by any progress
in the phenomenology of charged-particle fragmentation functions.
The quest for such progress has been increased dramatically by the advent of
new high-statistics data from HERA and LEP~1.
In particular, meaningful tests of the QCD-improved parton model can be
performed only beyond LO, so that knowledge of fragmentation functions
with NLO evolution is essential.
The present paper makes an effort to fill this gap.

By the factorization theorem, the use of our fragmentation functions is
not restricted to $e^+e^-$ collisions.
They may be applied to any other process by which charged pions and kaons
are produced inclusively.
However, one should bear in mind that our sets come with a certain range of
validity, which is specified in the Appendix.
For instance, the production of low-$p_T$ hadrons in high-energy
hadron-hadron and $ep$ scattering has a tendency to probe small values of
$x$ that lie outside this range.
Moreover, such hadrons are likely to originate from gluons, whose
fragmentation function is not yet constrained so well by $e^+e^-$ data.
This leaves room for further improvements in the future.

\bigskip
\centerline{\bf ACKNOWLEDGMENTS}
\smallskip\noindent
We thank J.Ph.~Guillet for clarifying comments regarding \cite{chiap}.

\begin{appendix}

\section{Parameterizations}

For the reader's convenience, we present here simple parameterizations of the
$x$ and $Q^2$ dependence of our fragmentation functions.
It is necessary to quote a range of validity for these parameterizations.
In the case of $Q^2$, they agree with the results obtained
by explicit evolution to better than 10\% if 5~GeV${}\le\sqrt{Q^2}\le91$~GeV.
As for $x$, we are experiencing technical limitations connected with the
stability of the numerical integration for $x\to0$ and $x\to1$.
Our results should be reliable in the interval $0.1\le x\le 0.9$.
The upper bound on $x$ may be ignored in practice,
since the fragmentation functions take very small values at $x=0.9$ and
approach zero for $x\to 1$.

As usual, we introduce the scaling variable
\be
\bar{s}=\ln{\frac{\ln{(Q^2/\Lambda^2)}}{\ln{(Q^2_0/\Lambda^2)}}}
\ee
and parameterize our results by simple functions in $x$ with coefficients
that are polynomials in $\bar{s}$.
We find that the template
\begin{equation}
\label{temp}
D(x,Q^2)=Nx^{\alpha}(1-x)^{\beta}(1+x)^{\gamma}
\end{equation}
works best in our case.
We now list the parameters to be inserted in (\ref{temp})
for charged pions and kaons both in LO and NLO.
At $Q^2=Q_0^2$, these parameterizations coincide with (\ref{pilo}--\ref{kanlo})
as they should.
\begin{enumerate}
\item LO fragmentation functions for $(\pi^++\pi^-)/2$:
\begin{itemize}
\item $D_V^{(\pi,LO)}(x,Q^2)$:
\begin{eqnarray}
      N&\n=\n&     0.551-0.053\bar{s}-0.032\bar{s}^2\nonumber\\
      \alpha&\n=\n& -1.0-0.0272\bar{s}\nonumber\\
      \beta&\n=\n&  1.2+0.67\bar{s}\nonumber\\
      \gamma&\n=\n& -0.393\bar{s}
\end{eqnarray}
\item $D_S^{(\pi,LO)}(x,Q^2)$:
\begin{eqnarray}
      N&\n=\n&     1.23+2.85\bar{s}-1.60\bar{s}^2\nonumber\\
      \alpha&\n=\n& -1.0+0.447\bar{s}-0.266\bar{s}^2\nonumber\\
      \beta&\n=\n&  4.77-2.88\bar{s}+2.05\bar{s}^2\nonumber\\
      \gamma&\n=\n& -9.01\bar{s}+4.36\bar{s}^2
\end{eqnarray}
\item $D_G^{(\pi,LO)}(x,Q^2)$:
\begin{eqnarray}
      N&\n=\n&      3.77-5.49\bar{s}+2.14\bar{s}^2\nonumber\\
      \alpha&\n=\n& -1.73\bar{s}+0.329\bar{s}^2\nonumber\\
      \beta&\n=\n&  2.0+1.00\bar{s}+0.44\bar{s}^2\nonumber\\
      \gamma&\n=\n& -2.05\bar{s}+2.48\bar{s}^2
\end{eqnarray}
\end{itemize}
\item NLO fragmentation functions for $(\pi^++\pi^-)/2$:
\begin{itemize}
\item $D_V^{(\pi,NLO)}(x,Q^2)$:
\begin{eqnarray}
      N&\n=\n&     0.338-0.064\bar{s}-0.0105\bar{s}^2\nonumber\\
      \alpha&\n=\n& -1.0-0.059\bar{s}\nonumber\\
      \beta&\n=\n&  1.2+0.60\bar{s}\nonumber\\
      \gamma&\n=\n& -0.163\bar{s}
\end{eqnarray}
\item $D_S^{(\pi,NLO)}(x,Q^2)$:
\begin{eqnarray}
      N&\n=\n&     1.19+4.20\bar{s}-2.86\bar{s}^2\nonumber\\
      \alpha&\n=\n& -1.0+0.757\bar{s}-0.537\bar{s}^2\nonumber\\
      \beta&\n=\n&  5.26-5.22\bar{s}+3.62\bar{s}^2\nonumber\\
      \gamma&\n=\n& -13.6\bar{s}+8.17\bar{s}^2
\end{eqnarray}
\item $D_G^{(\pi,NLO)}(x,Q^2)$:
\begin{eqnarray}
      N&\n=\n&     3.65-6.00\bar{s}+2.52\bar{s}^2\nonumber\\
      \alpha&\n=\n& -1.49\bar{s}-0.213\bar{s}^2\nonumber\\
      \beta&\n=\n&  2.0-0.23\bar{s}+1.52\bar{s}^2\nonumber\\
      \gamma&\n=\n& -6.53\bar{s}+8.43\bar{s}^2
\end{eqnarray}
\end{itemize}
\item LO fragmentation functions for $(K^++K^-)/2$:
\begin{itemize}
\item $D_V^{(K,LO)}(x,Q^2)$:
\begin{eqnarray}
      N&\n=\n&     0.155-0.010\bar{s}-0.010\bar{s}^2\nonumber\\
      \alpha&\n=\n& -1.0-0.012\bar{s}\nonumber\\
      \beta&\n=\n&  1.4+0.66\bar{s}\nonumber\\
      \gamma&\n=\n& -0.572\bar{s}
\end{eqnarray}
\item $D_S^{(K,LO)}(x,Q^2)$:
\begin{eqnarray}
      N&\n=\n&     0.216-0.0744\bar{s}\nonumber\\
      \alpha&\n=\n& -1.0-0.196\bar{s}\nonumber\\
      \beta&\n=\n&  1.69+0.69\bar{s}\nonumber\\
      \gamma&\n=\n& -0.195\bar{s}
\end{eqnarray}
\item $D_G^{(K,LO)}(x,Q^2)$:
\begin{eqnarray}
      N&\n=\n&     0.324-0.506\bar{s}+0.205\bar{s}^2\nonumber\\
      \alpha&\n=\n& -1.860\bar{s}\nonumber\\
      \beta&\n=\n&  2.0+1.67\bar{s}\nonumber\\
      \gamma&\n=\n& -0.235\bar{s}+2.68\bar{s}^2
\end{eqnarray}
\end{itemize}
\item NLO fragmentation functions for $(K^++K^-)/2$:
\begin{itemize}
\item $D_V^{(K,NLO)}(x,Q^2)$:
\begin{eqnarray}
      N&\n=\n&     0.099-0.0198\bar{s}-0.0027\bar{s}^2\nonumber\\
      \alpha&\n=\n& -1.0-0.060\bar{s}\nonumber\\
      \beta&\n=\n&  1.4+0.60\bar{s}\nonumber\\
      \gamma&\n=\n& -0.233\bar{s}
\end{eqnarray}
\item $D_S^{(K,NLO)}(x,Q^2)$:
\begin{eqnarray}
      N&\n=\n&     0.214-0.00073\bar{s}-0.0063\bar{s}^2\nonumber\\
      \alpha&\n=\n& -1.0-0.0067\bar{s}\nonumber\\
      \beta&\n=\n&  2.16+0.57\bar{s}\nonumber\\
      \gamma&\n=\n& -0.912\bar{s}
\end{eqnarray}
\item $D_G^{(K,NLO)}(x,Q^2)$:
\begin{eqnarray}
      N&\n=\n&      0.372-0.606\bar{s}+0.253\bar{s}^2\nonumber\\
      \alpha&\n=\n& -2.55\bar{s}+0.584\bar{s}^2\nonumber\\
      \beta&\n=\n&  2.0-0.52\bar{s}+1.82\bar{s}^2\nonumber\\
      \gamma&\n=\n& -7.43\bar{s}+9.24\bar{s}^2
\end{eqnarray}
\end{itemize}
\end{enumerate}

\end{appendix}

\newpage

\vskip-6cm

\begin{figure}

\centerline{\bf FIGURE CAPTIONS}

\caption{\protect\label{fig1}$x$ distributions of the NLO pion fragmentation
functions at $Q_0^2=2$~GeV$^2$\hskip5cm}

\vskip-.2cm

\caption{\protect\label{fig5}$x$ distributions of the NLO kaon fragmentation
functions at $Q_0^2=2$~GeV$^2$\hskip5cm}

\vskip-.2cm

\caption{\protect\label{fig2}Cross section of inclusive charged-pion
production at $\protect\sqrt{s}=29$~GeV as a function of $x$.
The LO (dashed line) and NLO fits (solid line) are compared with TPC data
\protect\cite{tpc2}}

\vskip-.2cm

\caption{\protect\label{fig6}Cross section of inclusive charged-kaon
production at $\protect\sqrt{s}=29$~GeV as a function of $x$.
The LO (dashed line) and NLO fits (solid line) are compared with TPC data
\protect\cite{tpc2}}

\vskip-.2cm

\caption{\protect\label{fig3}Same as in Fig.~\protect\ref{fig2} for
$\protect\sqrt{s}=5.2$~GeV and DASP data \protect\cite{dasp}\hskip5cm}

\vskip-.2cm

\caption{\protect\label{fig4}Same as in Fig.~\protect\ref{fig2} for
$\protect\sqrt{s}=10$~GeV and ARGUS data \protect\cite{argus}\hskip5cm}

\vskip-.2cm

\caption{\protect\label{fig7}Same as in Fig.~\protect\ref{fig6} for
$\protect\sqrt{s}=10$~GeV and ARGUS data \protect\cite{argus}\hskip5cm}

\vskip-.2cm

\caption{\protect\label{fig13}Cross section of inclusive charged-hadron
production at $\protect\sqrt{s}=91$~GeV as a function of $x$.
The dashed (solid) line corresponds to the prediction on the basis of
(\protect\ref{sighadc}) at LO (NLO)}

\vskip-.2cm

\caption{\protect\label{fig14}Ratio of the $x$ distributions of
inclusive charged-hadron production at $\protect\sqrt{s}=91$ and $29$~GeV.
The dashed (solid) line corresponds to the prediction on the basis of
(\protect\ref{sighadc}) at LO (NLO).
The points correspond to DELPHI data \protect\cite{delphisig}
divided by MARK~II data \protect\cite{mark2sig},
taking into account only the statistical errors of the latter}

\vskip-.2cm

\caption{\protect\label{fig10}Scaling violation in the cross section of
inclusive charged-hadron production.
Shown is the $Q^2$ dependence for
$x=0.075,0.15,0.25,0.35,0.5,0.75,0.9$.
The dashed (solid) line corresponds to the prediction on the basis of
(\protect\ref{sighadc}) to LO (NLO).
The data are taken from
\protect\cite{dasp,argus,delphisig,mark2sig,amysig,tassosig,cellosig}
and are partly interpolated in $x$}

\vskip-.2cm

\caption{\protect\label{figopal}Cross section of inclusive charged-hadron
production at $\protect\sqrt{s}=91$~GeV as a function of $x$.
The dashed (solid) line corresponds to the prediction on the basis of
(\protect\ref{sighadc}) at LO (NLO).
The data are taken by OPAL \protect\cite{opal} and represent the first
direct measurement of the gluon fragmentation function for charged
hadrons}

\end{figure}

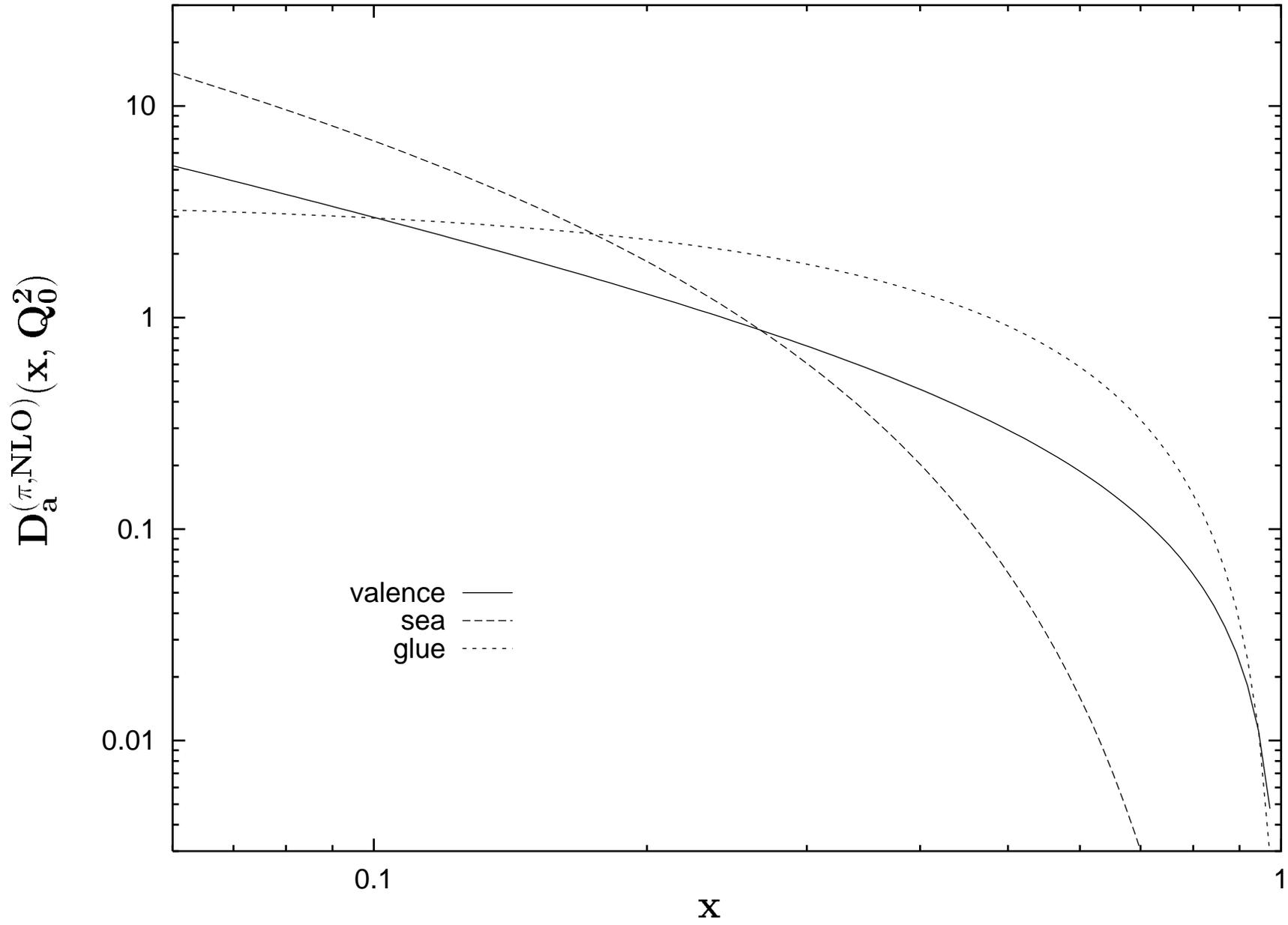
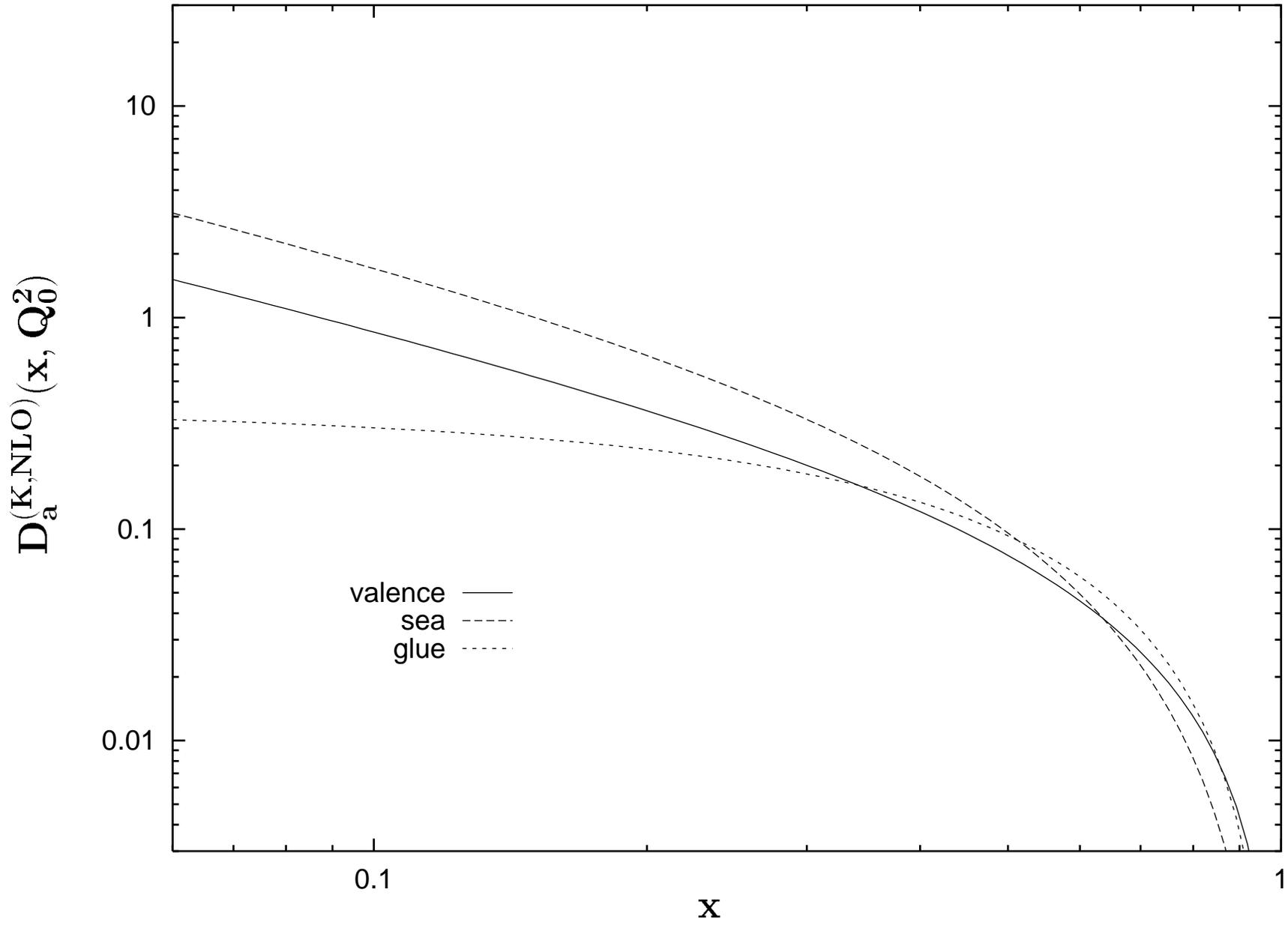
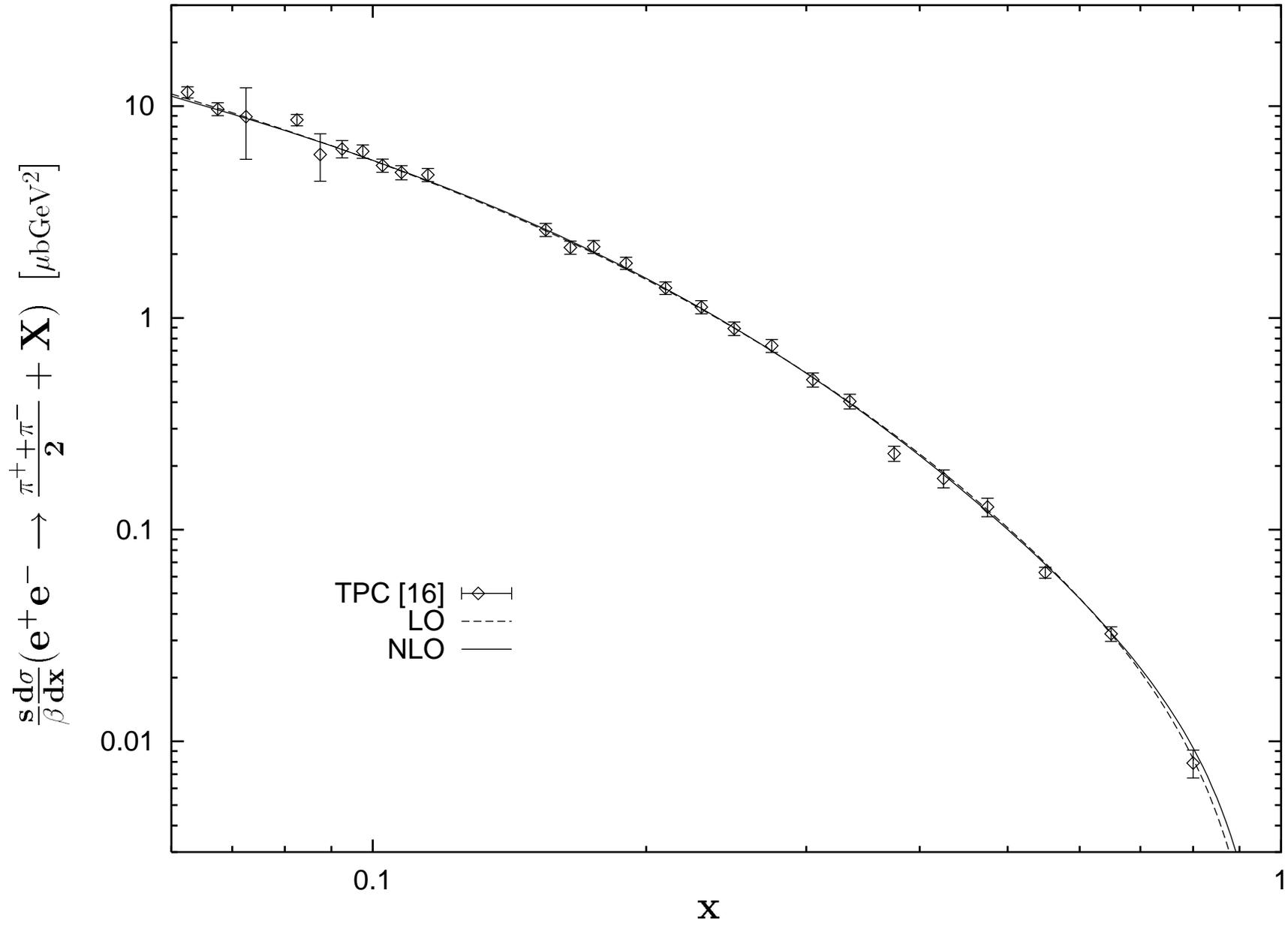
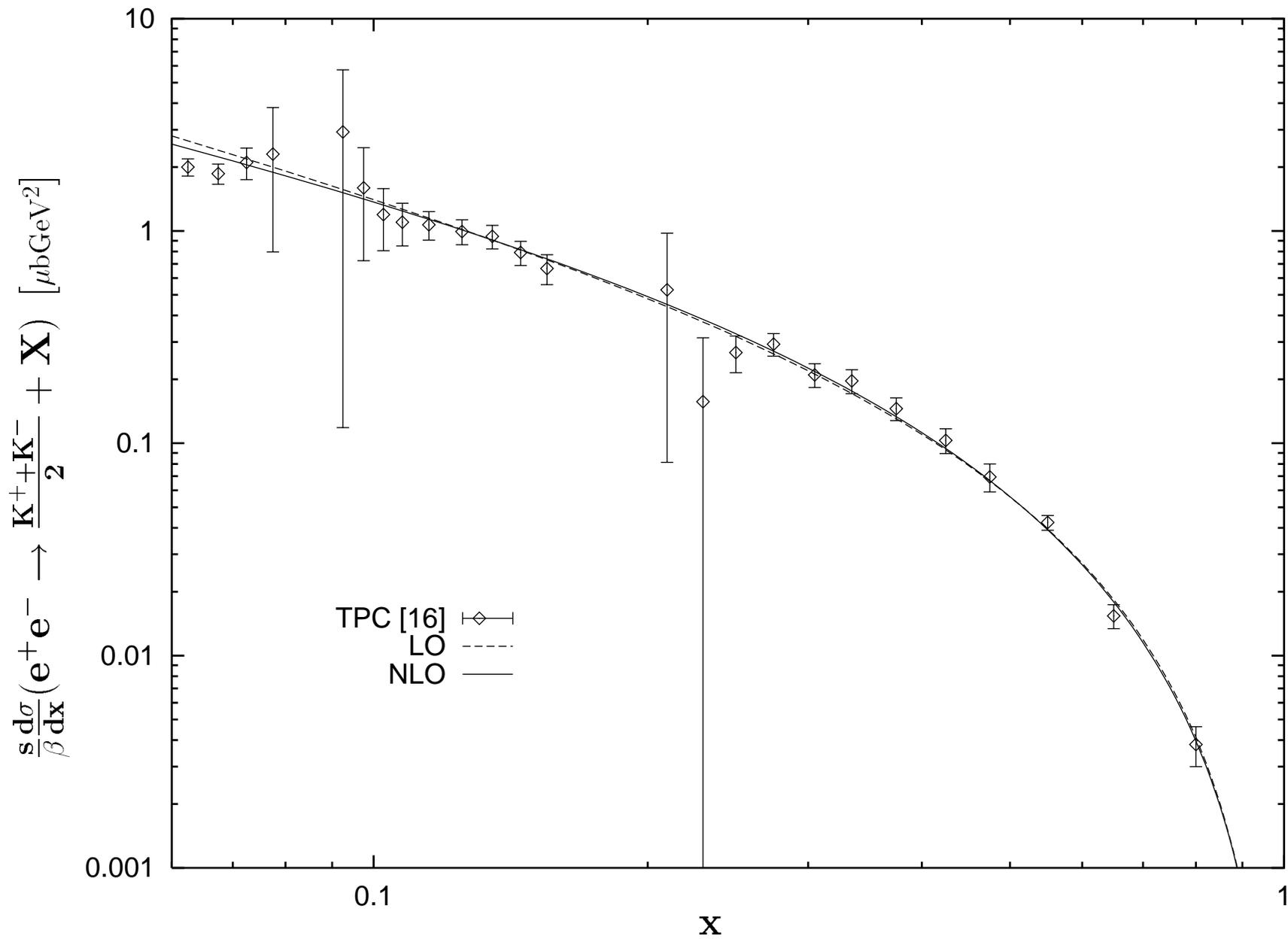
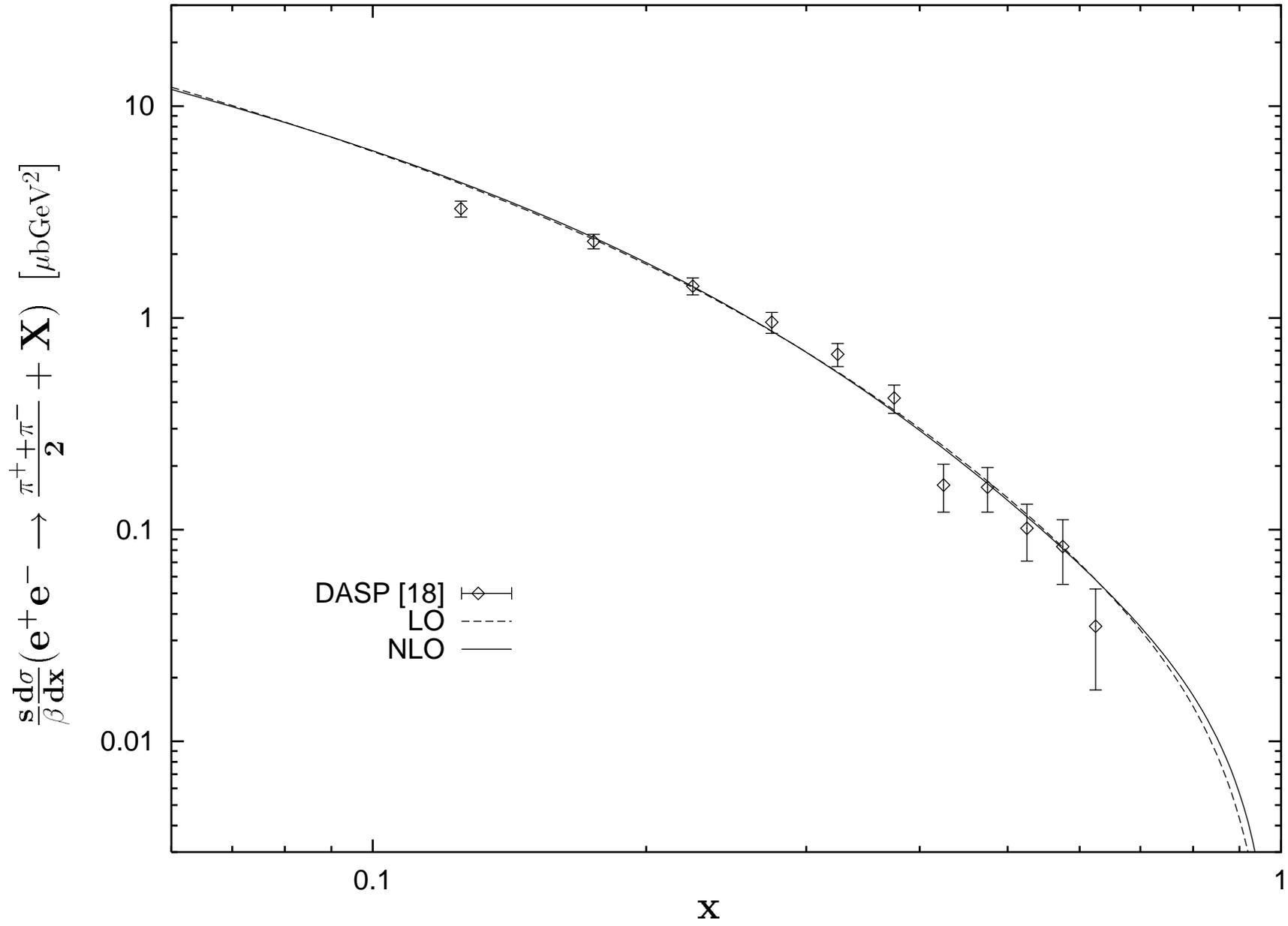
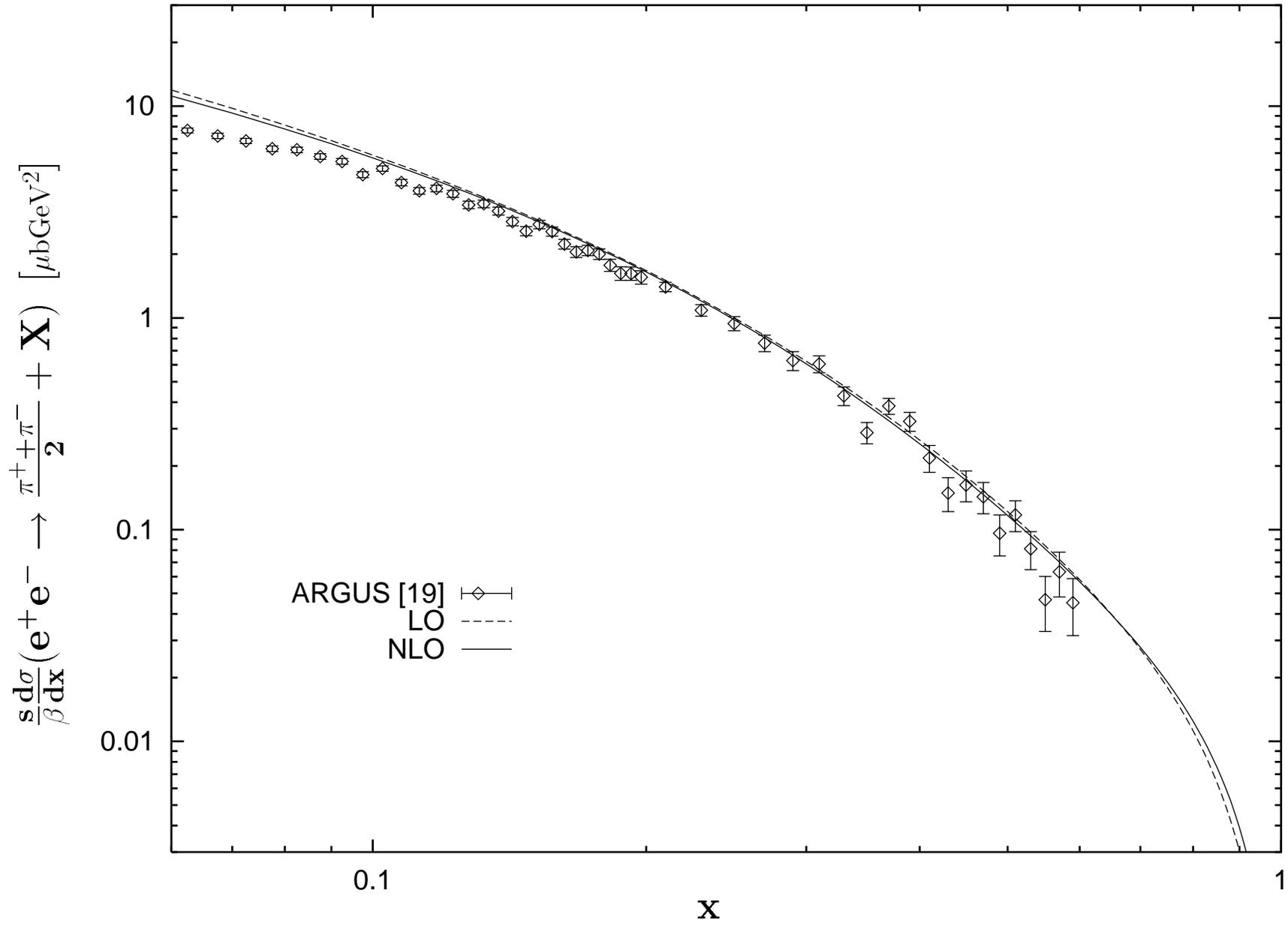
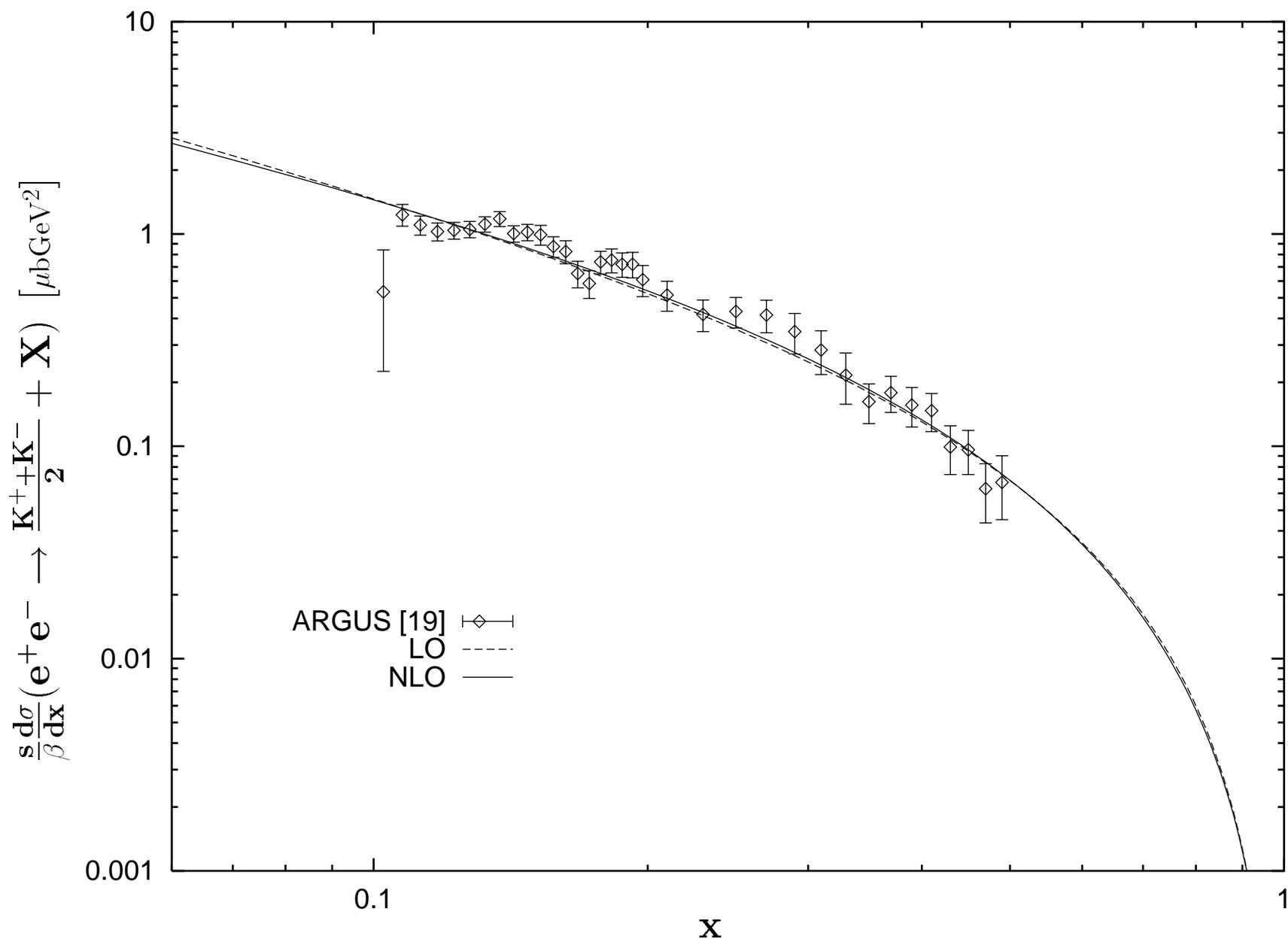
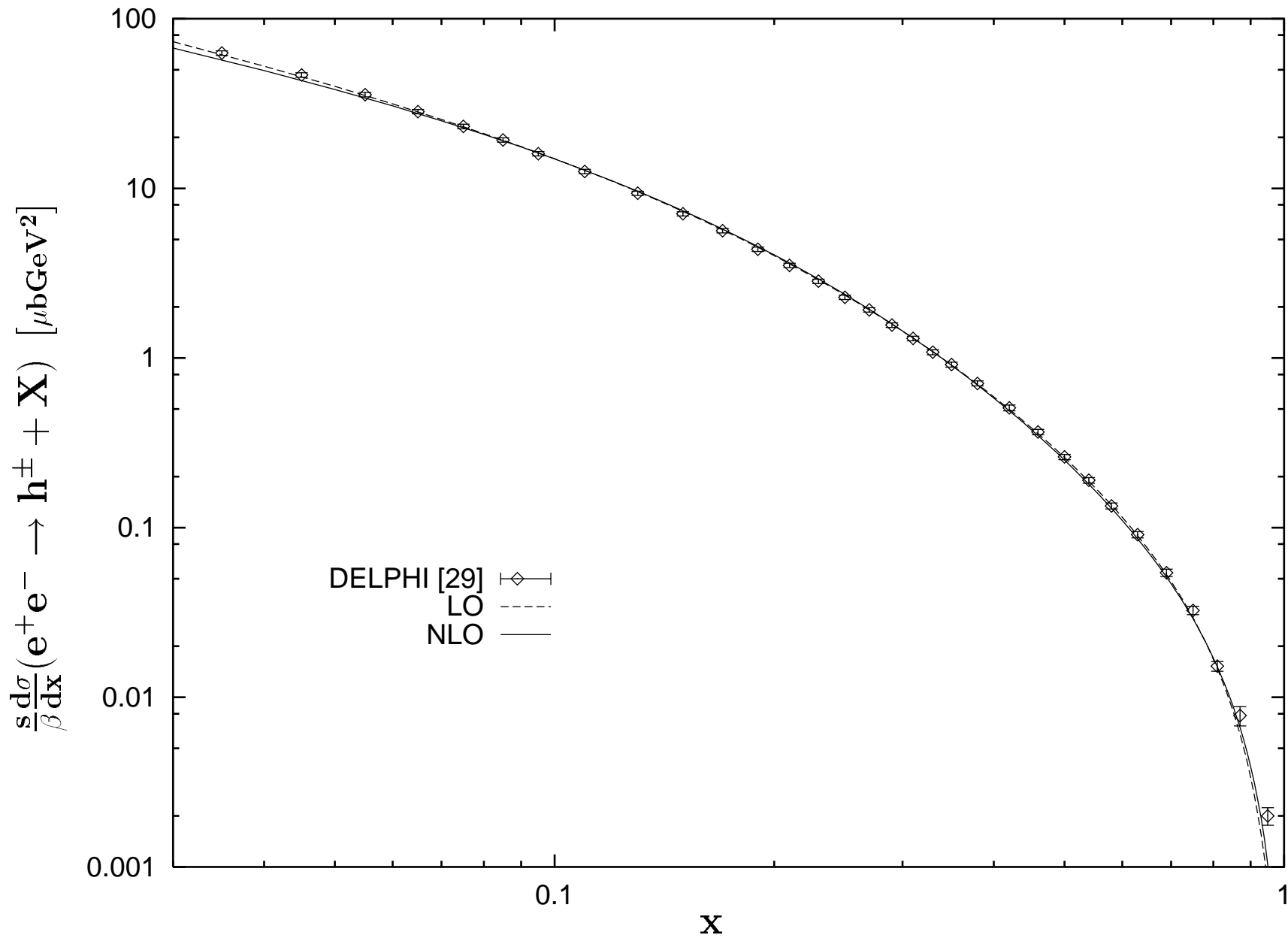
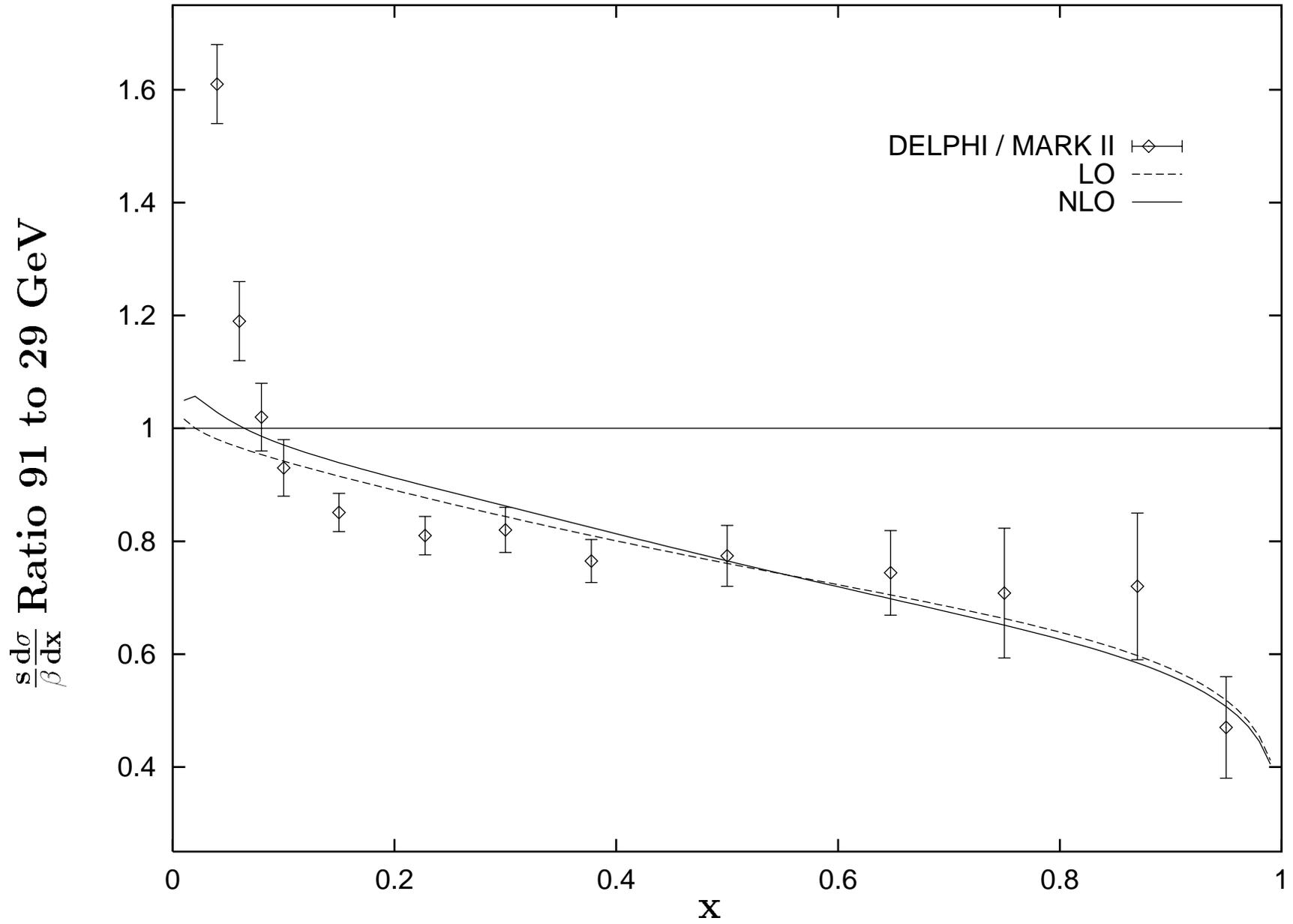
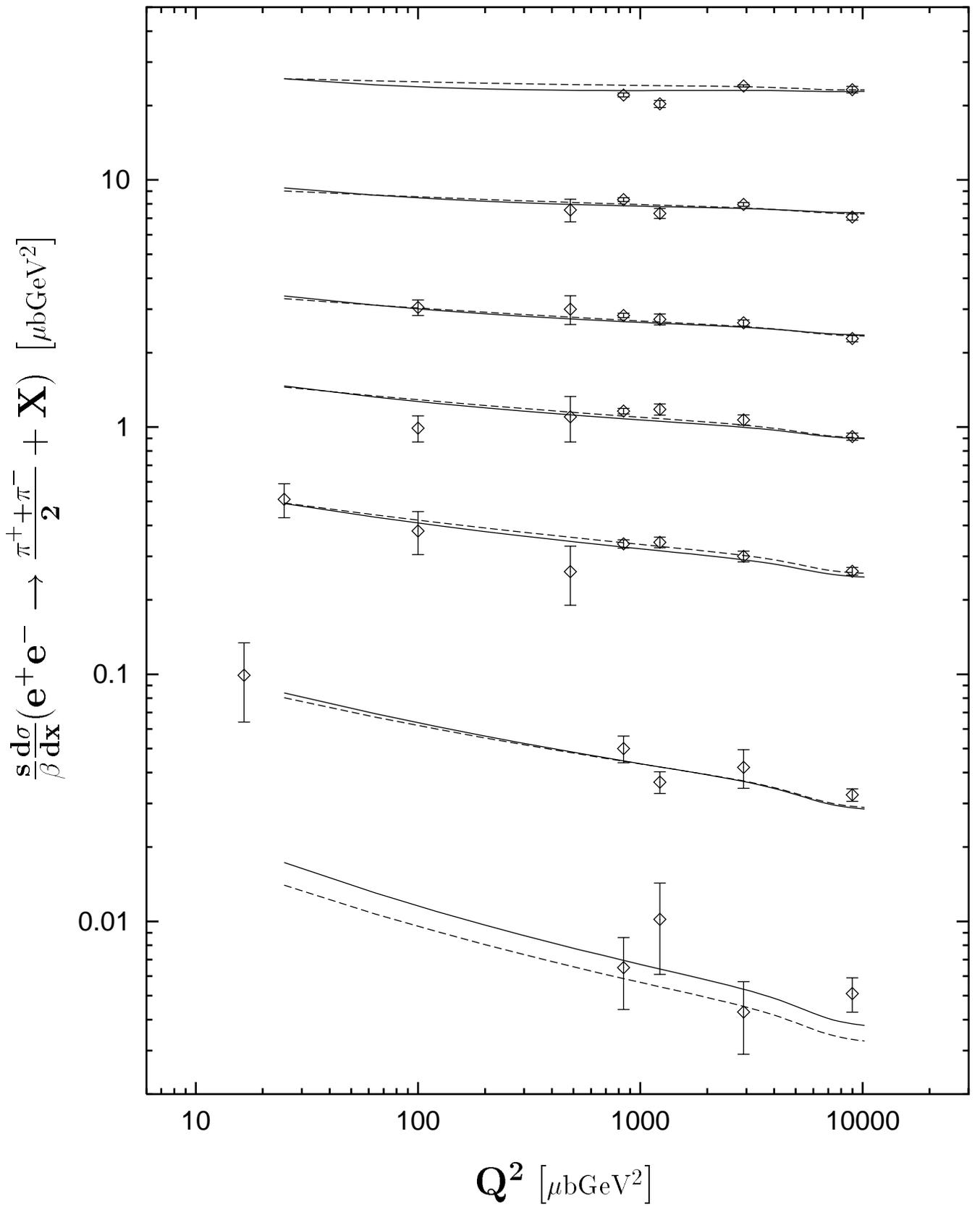
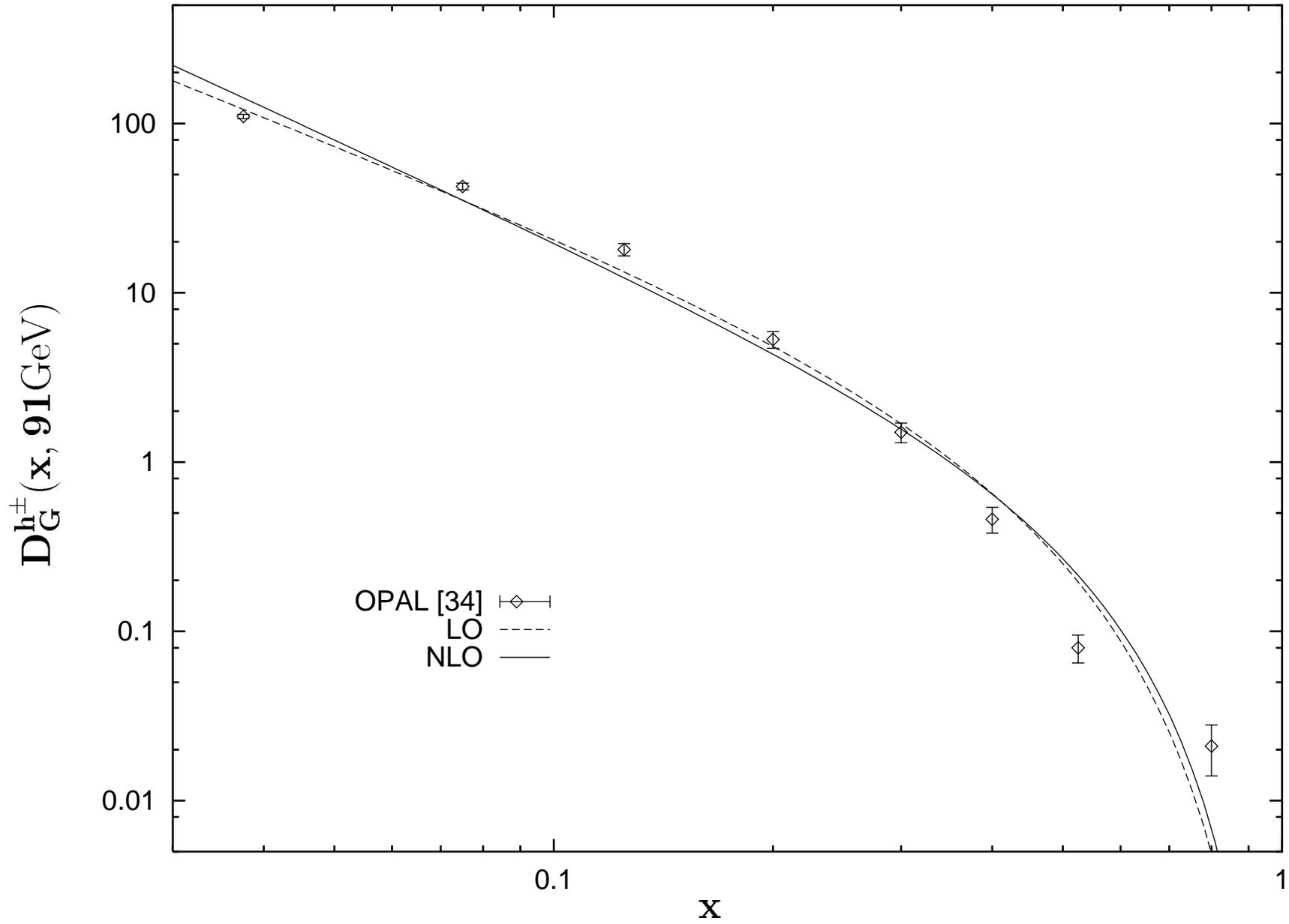
\end{document}